\documentstyle[aps,multicol,epsfig]{revtex}

\begin{document}
\title{Metal-insulator transition in PF$_6$ doped polypyrrole:
interchain charge transfer versus electronic correlation}
\author{H.C.F. Martens and H.B. Brom}
\address{Kamerlingh Onnes Laboratory, Leiden University, P.O. Box 9504,\\
2300 RA Leiden, The Netherlands}
\author{R. Menon}
\address{Department of Physics, Indian Institute of Science, Bangalore, 560012\\
India}
\date{\today}
\maketitle

\begin{abstract}We performed dielectric spectroscopy on polypyrrole near the metal-insulator
transition (MIT) down to 2~K. We evaluate the dependence of the plasma frequency
($\omega_p$) and the scattering time ($\tau$) on the distance to the MIT,
characterized by the Fermi-level relative to the band-edge of extended states,
$E_F-E_c$. Especially the strong decrease of $\tau$ with increasing $E_F-E_c$ is
in conflict with the usually applied models for the MIT. Although morphology and
disorder are important, the MIT is Mott-Hubbard-like being dominated by the
competition of interchain charge transfer and electronic correlation.
\end{abstract}


\begin{multicols}{2}
\settowidth{\columnwidth}{aaaaaaaaaaaaaaaaaaaaaaaaaaaaaaaaaaaaaaaaaaaaaaaaa}

In general the occurrence of a metal-insulator transition (MIT) depends on
several parameters\cite{Hirsch87}. In the absence of electron-electron
interactions  a random potential $W$ localizes charge carriers\cite{Anderson58}
due to the interference of multiply scattered electronic waves. For this process
the scaling-theory of Anderson localization\cite{Lee85} provides a convenient
framework. To describe electronic interactions, one usually introduces the
on-site interaction energy $U$ and inter-site interaction energy $V$ in the
so-called extended Hubbard hamiltonian\cite{Hubbard64}. Often the
electron-electron interactions suppress the conductivity, e.g. at half-filling
for $W=0$ and $V=0$, when $U$ becomes comparable to the inter-site
charge-transfer integral the so-called Mott-Hubbard metal-insulator transition
occurs. As reduced dimensionality enhances the effects of disorder and
interactions\cite{Lee85}, the geometry of the electronic system plays a profound
role as well. For instance, in a strictly one-dimensional system for any amount
of disorder all states are localized due to repeated
backscattering\cite{MottTwose61}. Moreover, even in the absence of disorder,
one-dimensional systems are insulating due to the Peierls instability and/or
finite interactions which, in one dimension, always induce a gap at the Fermi
level\cite{March96}. In doped conjugated polymers, it is widely recognized that
structural disorder drives the MIT\cite{Yoon94,Kohlman97,Lee95,Joo98}, while
recent observations of density wave dynamics points to the importance of
electronic correlations\cite{Lee00}. On a microscopic scale charge transport is
inevitably anisotropic: the charge transfer perpendicular to the chain is weak
compared to that along the chain, which enhances the effects of disorder and
electronic correlation. Nonetheless, a positive temperature coefficient of the
resistivity\cite{Yoon94,Hagiwara90} and a negative dielectric constant in the
microwave and far-infrared regime\cite{Kohlman97,Martens00PRB} indicate the
presence of a truly metallic state. In a previous study we pointed out a
correlation between the plasma frequency $\omega_p$ and scattering time $\tau$
which together characterize the free-carrier dynamics\cite{Martens00PRB}. To
explain the required finite density of delocalized states at the Fermi level
interchain charge transfer had to be sufficiently strong to overcome the
disorder- and interactions-induced localization\cite{Martens00PRB}.

In the present work we extend the dielectric measurements as a function of
frequency $\omega=2\pi f$ on PF$_6$ doped polypyrrole (PPy) around the MIT from
300~K to 2~K. The specific $T$- and $\omega$-dependent response enables to
characterize the position with respect to the MIT for both insulating and
metallic samples. We evaluate the dependence of $\omega_p$ and $\tau$ as a
function of this distance. Especially the strong decrease of $\tau$ when going
deeper into the metallic phase is in conflict with the usually applied models for
the MIT. This discrepancy provides a strong clue for the role of interchain
coupling and electronic correlation, indicating that the MIT may be considered as
a Mott-Hubbard transition. We suggest a mechanism that explains the influence of
structural disorder in this transition.

\begin{figure}
\leavevmode \hbox{\psfig{figure=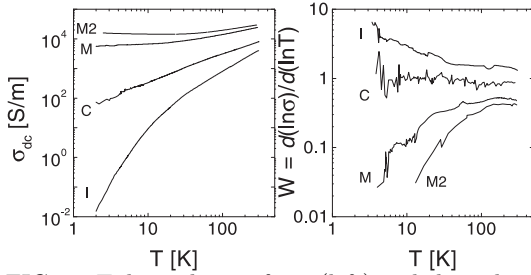,width=7cm}} \noindent{\caption{$T$
dependence of $\sigma_{\rm dc}$ (left) and the reduced activation energy
$W=d\ln(\sigma_{\rm dc})/d\ln(T)$ (right).}\label{dc}}
\end{figure}

Free-standing films of PPy-PF$_6$ are prepared by electrochemical
polymerization as described in detail elsewhere\cite{Yoon94,Lee95}. Films have
been polymerized at different temperatures ($-40^\circ{\rm C}<T_{{\rm
pol}}<20^\circ{\rm C}$). One sample ({\it D}) has been slightly dedoped
after the synthesis. The dc conductivity ($\sigma_{{\rm dc}}$) of several
samples is shown in Fig.~\ref{dc}. The plot of the reduced activation energy
$W=d\ln(\sigma_{{\rm dc}})/d\ln(T)$ versus $T$, Fig.~\ref{dc}b, demonstrates
that samples in the metallic ({\it M2, M}), critical ({\it C}), and insulating
({\it I}) regime of the MIT are obtained. Sample {\it M2} ($T_{\rm
pol}=-40^\circ$C) exhibits a conductivity minimum around 13~K, below which
$\partial\sigma_{{\rm dc}}/\partial T<0$ indicating truly metallic behavior.
However, for both {\it M} and {\it M2}, $\partial\sigma_{{\rm dc}}/\partial
T\geq0$ at higher $T$ which shows that PPy-PF$_6$ is at best a ``bad'' metal.
For increasing $T_{{\rm pol}}$ the conductive properties deteriorate and the
system goes through the MIT. This can be attributed to an increase of the
structural disorder\cite{Yoon94}.
The electrodynamic properties have been studied by means of transmission
experiments\cite{Reedijk00Exp} in the frequency range 8--700~GHz
(0.27--23~cm$^{-1}$, 0.033--2.9~meV), which overlaps with both the microwave and
far-infrared regime. Both amplitude and phase are obtained by means of an ABmm
vector network-analyzer. The data are fitted to first principle transmission
formulae to derive the complex conductivity $\sigma ^{*}(\omega )=\sigma (\omega
)+i\omega \varepsilon _{0}\varepsilon (\omega )$, without the use of
Kramers-Kronig manipulation\cite{Reedijk00Exp}. Uncertainty in $\sigma$ and
$\varepsilon$ is less than 5\% in the range 100--600~GHz at all $T$.
\begin{figure}
\leavevmode \hbox{\psfig{figure=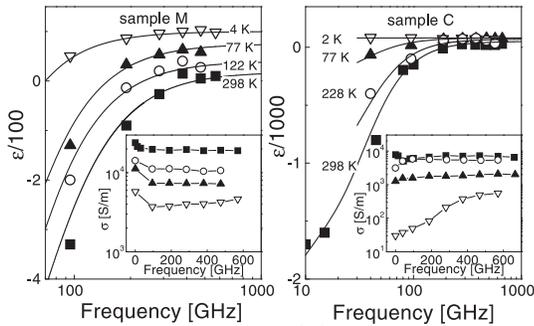,width=7cm}} \noindent{\caption{$T$
dependence of $\varepsilon(\omega)$ of {\it M} and {\it C}. Solid lines are fits
to the Drude model. The insets show $\sigma(\omega)$ of {\it M} and {\it C} at
the same $T$. Errorbars are less than the symbol size.}\label{epstdep}}
\end{figure}
The $T$-dependent $\varepsilon(\omega)$ of {\it C} and {\it M} are
shown in Fig.~\ref{epstdep}.
For {\it C} $\varepsilon<0$ only at high $T$, while for {\it M} a
negative contribution to $\varepsilon(\omega)$ persists
down to low $T$. The negative $\varepsilon$
agrees with the findings of Kohlman and co-workers\cite{Kohlman97}, and
establishes the presence of free carriers in PPy-PF$_6$.
Solid lines are fits to the Drude model:
\begin{equation}
\varepsilon(\omega)=\varepsilon_b -
{{\omega_p^2\tau^2}\over{1+\omega^2\tau^2}},\label{Drude}
\end{equation}
with $\tau$ the free-carrier scattering time, $\varepsilon_b$ the background
dielectric constant due to polarization of the lattice and localized carriers,
and $\omega_p$ the plasma frequency:
\begin{equation}
\omega_p=\sqrt{n_fe^2\over\varepsilon_0m^*}\label{wp},
\end{equation}
with $n_f$ the free-carrier density, $e$ electronic charge, $m^*$ the effective
mass, and $\varepsilon_0$ the vacuum permittivity. The insets of
Fig.~\ref{epstdep} show $\sigma(\omega)$ for {\it C} and {\it M} at the same
$T$'s. Note that for {\it M}, $\partial\sigma/\partial\omega<0$ at low $\omega$
at all $T$ as expected for free carriers, while for {\it C} only at room
temperature $\partial\sigma/\partial\omega<0$ is observed.

\begin{figure}
\leavevmode \hbox{\psfig{figure=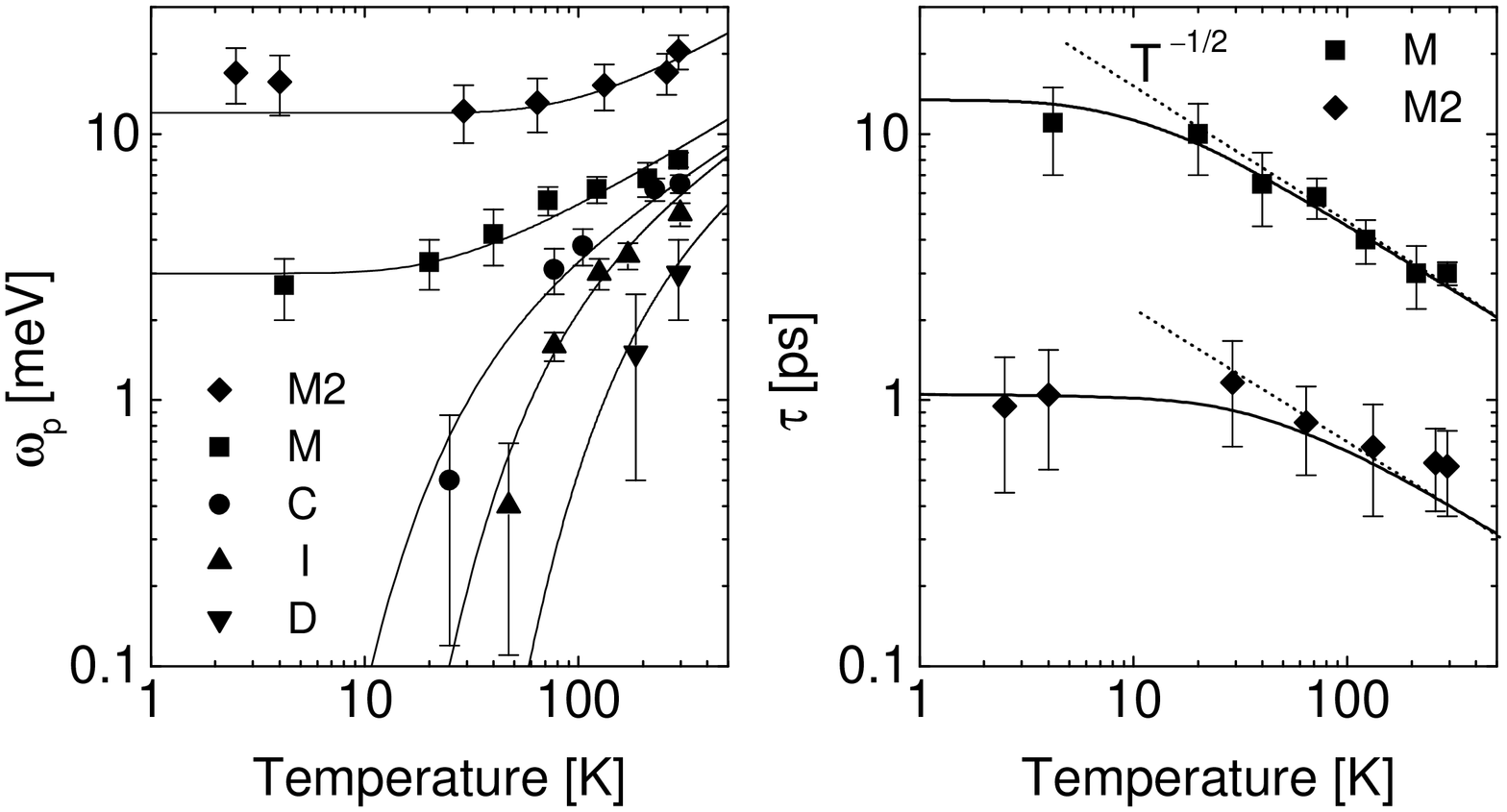,height=7cm,angle=-90}}
\noindent{\caption{$\omega_p$(left) and $\tau$ (right) as a function of $T$.
Solid lines are explained in the text; dotted lines indicate $\tau\propto
T^{-1/2}$.}\label{wptaut}}
\end{figure}
By fitting to Eq.~(\ref{Drude}), the $\varepsilon(\omega,T)$ data lead to
$\omega_p(T)$ as shown in Fig.~\ref{wptaut}a. The $T$-dependent data do not extend
to low enough $\omega$ to derive $\tau$ from $\varepsilon(\omega)$. However,
according to Drude $\sigma(\omega=0)=\omega_p^2\tau\varepsilon_0$, which is
estimated by $\sigma_{\rm dc}-\sigma(95\rm\ GHz)$\cite{Martens00PRB}. For metallic
{\it M} and {\it M2}, $\tau(T)$ could thus be obtained, see Fig.~\ref{wptaut}b.

\begin{figure}
\leavevmode \hbox{\psfig{figure=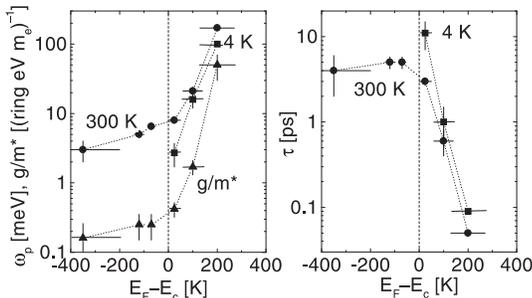,width=7cm}}
\noindent{\caption{Dependence of $\omega_p$, $g/m^*$, and $\tau$ on $(E_F-E_c)$
at $T=300$~K and $T=4$~K. Also included are PPy-PF$_6$ data taken from
Ref.~\protect\cite{Kohlman97}. Dotted lines are guides to the eye. Errorbars are
indicated, or less than the symbol size.}\label{wptEfEc}}
\end{figure}

First we consider the behavior of $\omega_p(T)$ near the MIT. For {\it M2} and
{\it M}, $\omega_p$ decreases when cooling but remains finite at low $T$: for
these metallic samples free carriers are present at $T=0$. For {\it C}, {\it I},
and {\it D}, $\omega_p$ strongly drops when lowering $T$ which shows that the
free carrier-density vanishes as $T\rightarrow0$. In comparison with conventional
metals where $\omega_p\sim$~1--10~eV, the $\omega_p$'s in PPy-PF$_6$ are very
low. For these systems near the MIT only a fraction of the carriers occupies
extended states (0.1\% if $m^*$ equals the free electron mass $m_e$; a larger
$m^*$ would imply a larger $n_f$) with the majority of the carriers residing in
localized states\cite{Kohlman97,Martens00PRB}. The observed increase of
$\omega_p$ with $T$ of both the metallic and insulating samples
shows that barely localized carriers can be thermally activated to higher-lying
extended states. For given Fermi-level $E_F$ and energetic boundary $E_c$ between
localized and extended states, the density of occupied extended states follows
from
\begin{equation}
n_f(T)=\int^\infty_{E_c} f_{{\rm FD}}(T)g(E)dE,\label{nfree}
\end{equation}
where $f_{{\rm FD}}$ denotes the Fermi-Dirac distribution function, $g(E)$ is the
density of states, and the Fermi level is defined as the zero of energy. Using
Eqs.~(\ref{wp}) and (\ref{nfree}) and assuming that $m^*$ is independent of $T$
and $g$ independent of $E$, the $\omega_p(T)$ data sets for each sample
can be excellently reproduced (solid lines in Fig.~\ref{wptaut}a) with
only two free parameters: $E_F-E_c$ and $g/m^*$.
The $T$-dependence of $\omega_p$ is fixed by
$E_F-E_c$; the absolute value is fixed by both.
For samples {\it M2}, {\it M},
{\it C}, {\it I}, and {\it D} $E_F-E_c$ is found to be: $100\pm 50$~K, $25\pm
10$~K, $-70\pm15$~K, $-120\pm20$~K, and $-350\pm150$~K respectively; values for
$g/m^*$ are presented in Fig.~\ref{wptEfEc}. Analysis of the PPy-PF$_6$ data
given in Ref.~\cite{Kohlman97} gives $E_F-E_c=200\pm100$~K. As expected, for
metallic samples $E_F>E_c$, while for {\it C}, {\it I}, and {\it D}, $E_F<E_c$.
While previous studies resorted to the so-called resistivity-ratio to
characterize conducting polymers\cite{Handbook}, the above results provide an
experimental characterization of the conductive state with respect to the MIT in
terms of a physical quantity. To our knowledge, for conducting polymers a
comparable quantitative analysis has not been reported, and, as demonstrated
below, provides a new perspective on the mechanisms behind the MIT.

Let us now turn to the derived $\tau$'s, see Fig.~\ref{wptaut}b. In view of
commonly reported $\tau\sim10^{-14}\rm\ s$ values in normal metals, the
scattering times in the disordered PPy-PF$_6$ are extremely long. Following
Ref.~\cite{Kohlman97}, if all carriers were delocalized in a three-dimensional
band, and hence $v\approx\sqrt{2E_F/m^*} \approx 5 \times 10^5 {\rm m/s}$, these
long $\tau$'s lead to anomalously long mean free-paths $\ell\approx0.2\ \mu$m and
$2\ \mu$m for {\it M2} and {\it M} respectively. However, only above $E_c$
carriers are mobile, hence the free-carrier velocity does not depend on the total
carrier density but only on the density of free carriers above
$E_c$\cite{Castner00}. If $k_BT<(E_F-E_c)$ then $v\approx\sqrt{(E_F-E_c)/m^*}$,
while $v\approx\sqrt{k_BT/m^*}$ for $k_BT>(E_F-E_c)$. The dotted lines in
Fig.~\ref{wptaut}b indicate $\tau\propto1/\sqrt{T}$ for {\it M} for $T>20\rm\ K$.
Since in this range $k_BT>(E_F-E_c)$ and $v\propto\sqrt{T}$, this suggests that
$\ell$ is $T$-independent. The increase of $\tau$ when cooling could indicate an
electron-phonon scattering mechanism. However, the $T$-independence of $\ell$ does not
support such a scenario, but indicates that scattering is
dominated by static disorder. Assuming constant $\ell$ and taking into account
the $T$-dependence of the carrier velocity expressed by $v=n_f^{-1}\int_{E_c}^\infty
f_{\rm FD}g\sqrt{(E-E_c)/m^*}dE$, $\tau(T)$ can be excellently reproduced
using the $(E_F-E_c)$ derived from $\omega_p(T)$ ($E_F-E_c$ fixes
the $T$-dependence of $\tau$). For {\it M2} due to
the errorbars we can not be conclusive on the $T$-dependence of $\tau$. Taking
$m^*=m_e$, $\ell=\rm 20\ nm$ and $\ell=\rm200\ nm$ for {\it M2} and
{\it M} respectively; larger $m^*$ would give shorter $\ell$. The reason for
$\ell$ to appear longer for the ``less metallic'' {\it M} is discussed below.

Fig.~\ref{wptEfEc} shows the derived values of $\omega_p$ and $\tau$ versus
$E_F-E_c$ at 300~K and 4~K and the ratio $g/m^*$ versus $E_F-E_c$, and is the
main result of our study. First, consider the dependence of $\omega_p$ on
$E_F-E_c$. As discussed above, the finite $\omega_p$ for $T=300$~K in the
insulating regime ($E_F<E_c$) reflects the thermal activation of barely localized
carriers to higher-lying extended states. In the metallic regime for
$T\rightarrow0$ the density of free-carriers $n_f\approx g\times(E_F-E_c)$ and,
for constant $g/m^*$, $\omega_p$ would increase according to
$\omega_p\propto\sqrt{E_F-E_c}$. However, as $g/m^*$ increases strongly in the
metallic regime, a much stronger increase of $\omega_p$ results.

The dependence of $\tau$ on $E_F-E_c$ near the MIT seems peculiar: when coming
from the insulating regime and crossing the MIT $\tau$ drops sharply. This
behavior persists in both the degenerate, $(E_F-E_c)>k_BT$, and non-degenerate,
$(E_F-E_c)<k_BT$, regimes. We note that this decrease of $\tau$ with $E_F-E_c$
agrees with a recently pointed out empirical correlation
$\tau\propto\omega_p^{-x}$ ($x\sim1.3$) over two orders of magnitude
$\omega_p$-range\cite{Martens00PRB}. According to the scaling theory of
Anderson-localization\cite{Lee85}, $E_F-E_c$ increases when the amplitude of the
disorder potential decreases, and $\ell\propto(E_F-E_c)^p$ with
$p>0.5$\cite{Castner00}, becomes larger. Hence, Anderson theory gives
$\tau=\ell/v\propto(E_F-E_c)^{p-1/2}$ which is a monotonically increasing
function for $E_F>E_c$, and is therefore in conflict with our experimental
observations, see Fig.~\ref{wptEfEc}b. It has been proposed that the MIT in
conducting polymers is better viewed in terms of percolation of metallic
islands\cite{Kohlman97,Joo98}. Then $E_F-E_c$ characterizes the barrier heights
between the metallic domains, and $E_F=E_c$ is the percolation threshold. In a
percolating metallic network $\tau$ corresponds to either the intrinsic
free-carrier scattering time in the metallic islands, or reflects the carrier
scattering due to finite-size effects of the metallic network. In the first case
$\tau$ should be independent of $E_F-E_c$, in the latter case $\tau$ should
increase with increasing $E_F-E_c$. Clearly, also this model for the MIT in
conducting polymers can not account for our experimental results. Hence, the
decrease of $\tau$ in the metallic regime can not be understood in terms of the
schemes usually applied to the MIT in conducting
polymers\cite{Yoon94,Kohlman97,Lee95,Joo98}, which only consider effects of
disorder. However, even in the absence of disorder, to obtain a {\em macroscopically}
conducting state interchain charge-transfer $t_c$ is a prerequisite: in the limit
of vanishing $t_c$ carriers are bound to individual polymer-chains giving rise to
an insulating state. Hence a poor $t_c$ will impede the formation of
a metallic state and determine the low-frequeny
conductive properties (only for $\hbar\omega>t_c$ the intrachain
transport dominates).

To explain the effect of $t_c$ on the charge transport properties of conducting
polymers, a tight-binding picture is illustrative. The bandwidth for interchain
transport is directly proportional to $t_c$, while the effective mass for
interchain transport is inversely proportional to $t_c$. However, for small $t_c$
this band becomes unstable for electronic interactions. Since optimally doped
PPy-PF$_6$ is quarter-filled, the inter-site interaction $V$ will be most
important. When $t_c$ lies below $V$ the interchain transfer can not overcome the
Coulomb repulsion and the carriers remain localized to the chains: the
system is an insulator. For $t_c>V$ wavefunctions are delocalized over adjacent
chains and the system becomes a metal. Such a mechanism has been demonstrated to
drive the MIT e.g. in quasi-one-dimensional, crystalline Bechgaard salts; only
for $t_c$ exceeding the correlation energy a three-dimensional metallic state is
formed\cite{Vescoli98}. The analogy goes even further, as in these Bechgaard
salts the metallic state is characterized by a small $\omega_p$ (spectral weight
1\%) as well\cite{Vescoli98}. The essential ingredient for this transition, a
small value of $t_c$ compared to the electron-electron interaction energy, is
also fulfilled in conducting polymers as illustrated by the following simple
calculation. The interchain charge transfer is typically
$t_c\sim$~0.01--0.1~eV\cite{Handbook,Vescoli98,Paulsson99} and $V =
e^2/(4\pi\varepsilon_0\varepsilon_b r)$, with $r$ the separation between charges.
For optimal doping of 1 charge per 4 monomers $r$ is a few nm and
$V\approx$~0.1~eV ($U$ will be even larger). Since $t_c$ and $V$ are of the same
order of magnitude the competetion between interchain charge transfer and
electron-electron correlations may play a dominant role in the occurence of the
MIT in conducting polymers. These conclusions are in agreement with a recent
numerical finding of a strong dependence of the MIT on interchain interactions in
polyacetylene\cite{Paulsson99}.

What should be expected for the carrier dynamics near the MIT in case of a
Mott-Hubbard transition? Starting in the metallic regime, when decreasing $t_c$,
the effective mass increases and it has even been suggested that $m^*$ diverges
near the critical point\cite{March96}, while the density of delocalized states
vanishes near the MIT\cite{Hubbard64,March96}. This implies a strong increase of
$g/m^*$ with increasing $E_F-E_c$ in qualitatitive agreement with our
experimental findings, see Fig.~\ref{wptEfEc}. To our knowledge, a relation
between $g/m^*$ and distance to the MIT is not available. Note that, given the
values $g\sim$~1--10~states/(eV ring) reported in
literature\cite{Kohlman97,Handbook,Raghunathan98}, Fig.~\ref{wptEfEc} suggests
$m^*$ to be in the 1--10$\times m_e$ range. The increase of $m^*$ in the
Mott-Hubbard scenario leads to a strongly decreasing free-carrier velocity $v\propto
\sqrt{(E_F-E_c)/m^*}$ near the critical point. Since the scattering probability
is proportional to the density of states available for scattering, i.e.
$g$\cite{Martens00PRB}, we expect $\tau\propto
g^{-1}\sqrt{m^*/(E_F-E_c)}$, which decreases for increasing $E_F-E_c$.
Hence the unusual decrease of $\tau$, and seemingly shorter
$\ell$, in the metallic regime supports our arguments that the MIT is driven by
the interplay between $t_c$ and $V$ which gives decreasing $g$ and increasing
$m^*$ near the critical regime.

How can we reconcile this picture with the observations that structural
disorder\cite{Yoon94,Kohlman97} drives the MIT? The interchain $\pi$-electronic
overlap $t_c$ depends {\em exponentially} on separation and orientation of
adjacent chain(segment)s and is therefore particularly sensitive to the local
ordering of polymer chains. Deviations from the optimal packing make the allowed
bandwidth exponentially narrow and the effective mass exponentially large. Due to
this extreme sensitivity, the increase of $t_c$ with respect to $U,V$ may be the
dominant driving mechanism for the MIT instead of disorder-induced localization.
Also X-ray diffraction studies\cite{Nogami94} demonstrate that a high
conductivity in PPy requires a close packing of adjacent pyrrole rings which
should be favorable for the $\pi$-electronic overlap. The dependence of the MIT
on pressure, which enhances $t_c$\cite{Vescoli98,Vakiparta93}, confirms this
point of view.

In summary, we have studied the free-carrier dynamics in PPy-PF$_6$ in the
vicinity of the MIT. The specific $T$-dependence allows to characterize the
position of the samples with respect to the MIT. It is demonstrated that
$\omega_p$ increases, and, unexpectedly that $\tau$ strongly drops in better
conducting samples. These results are inconsistent with models which only
consider the effect of disorder-induced localization. Instead, we have
argued that the behavior in the vicinity of the MIT can be explained
in terms of a Mott-Hubbard transition due to the competition between
electronic correlation and interchain charge transfer.
Structural disorder could induce such a
Mott-Hubbard transition via the strong sensitivity of $t_c$ on the packing of
individual chains.

This work is part of the research program of FOM.

\end{multicols}

\end{document}